\documentclass[onecollarge]{svjour2}       
\usepackage{graphicx}
\usepackage{alltt}
\begin{document}

\title{Earth tides in MacDonald's model}
\author{Sylvio Ferraz-Mello}
\institute{Instituto de Astronomia 
Geof\'{\i}sica e Ci\^encias Atmosf\'ericas\\Universidade de S\~ao Paulo, Brasil \\
sylvio [at] usp.br}
\titlerunning{Synchronization}

\maketitle

\begin{abstract}
We expand the equations used in MacDonald's 1964 theory and Fourier analyze the tidal variations of the height at one point on the Earth surface, or, alternatively, the tidal potential at such point. It is shown that no intrinsic law is relating the lag of the tide components to their frequencies. In other words, no simple rheology is intrinsically fixed by MacDonald's equations. The same is true of the modification proposed by Singer(1968). At variance with these two cases, the modification proposed by Williams and Efroimsky (2012) fix the standard Darwin rheology in which the lags are proportional to the frequencies and their model is, in this sense, equivalent to Mignard's 1979 formulation of Darwin's theory.
\end{abstract}

\def\beq{\begin{equation}}
\def\endeq{\end{equation}}
\def\begdi{\begin{displaymath}}
\def\enddi{\end{displaymath}}
\def\ep{\varepsilon}
\def\epp{\varepsilon^\prime}
\def\eppp{\varepsilon^{\prime\prime}}
\def\App{A^\prime}
\def\Appp{A^{\prime\prime}}
\def\bpp{\beta^\prime}
\def\bppp{\beta^{\prime\prime}}
\def\Cpp{C^\prime}
\def\Cppp{C^{\prime\prime}}
\def\CppQ{C^{\prime 2}}
\def\CpppQ{C^{\prime\prime 2}}
\def\aprmaior{\;\buildrel\hbox{\geq}\over{\sim}\;}    
\def\defeq{\;\buildrel\hbox{\small def}\over{\,=}\;}    
\def\speq{\hspace{1mm} = \hspace{1mm}}    
\def\FRH{Ferraz-Mello et al (2008)}

\keywords{MacDonald theory - Singer theory - Darwin rheology - tide lags - Mignard's hypothesis}
\section{Introduction}
MacDonald's name is often used to designate tidal theories in which the tide lags are frequency independent (see e.g. Ferraz-Mello et al. 2008 footnotes 4 and 6). This is however wrong. 
Indeed MacDonald (1964) adopted a constant lag $\delta$ and integrated the evolution equations keeping this value unchanged during the whole evolution. However, what is called lag in MacDonald's theory is not the same quantity as adopted in Kaula(1964) and/or in modern theories. Using Kaula's notations (see Williams and Efroimsky, 2012), frequency independent lags actually mean that all $\ep_{lmpq}$ are equal and independent of the frequencies $\chi_{lmpq}$. In Ferraz-Mello et al.'s notations, it means 
$\ep_0=\ep_1=\ep_2=\cdots$. This does not happen in MacDonald's tide and thus it is not correct to associate Mac Donald's name to such tidal theories.

We investigate in this note how lag the tide components in MacDonald's model. First we have to understand what is done in that paper. The key equations are MacDonald's eqns.(20$-$21). Eqn. (20) is introduced as follows:

\begin{quotation}\noindent Let us suppose that at the earth's surface the tide lags by an angle $\delta$ , so that, at a time
$f$ (the moon's angular position, $f$ , is used to keep time), the potential is not ${\cal{W}}(f)$ but rather
${\cal{W}}(f-\delta)$. Thus, at a time $f$ , the extra potential due to the tidal deformation of the surface
of the earth is:
\begdi
{\cal{W}}=k_2 {\cal{W}}_2(f-\delta)
\enddi
\vspace*{-1.3cm}
\begin{flushright}
(MD. eqn. 20)
\end{flushright}
\vspace*{0.2cm}\noindent
since the rotation carries the bulge forward and \underline {the zenith distance becomes $\delta$}
\end{quotation}
(We omit the terms with subscripts $3,4,\dots,n$, because they are irrelevant in this discussion).

This quotation is the point where ambiguity is introduced in MacDonald's paper because $f$ is the true anomaly, which is not a uniform quantity, and which could not be used as equivalent to time.  

The only way to overcome this ambiguity is learning from MacDonald's equations what he actually did. This is done in Sections 2 and 3 of this note. The results of these two sections are founded on MacDonald's eqn. (21), which is given in his paper as follows:

\begin{quotation}
\noindent The external potential at a time $f$ is
\begdi
{\cal{U}}=  \frac{GmR_\oplus^5}{r_f^3 r_{f-\delta}^3} k_2 P_2(\cos \delta)
\enddi
\vspace*{-1.3cm}
\begin{flushright}
(MD. eqn. 21)
\end{flushright}
\vspace*{0.5cm}\noindent
where $r_f$ corresponds to the instantaneous distance to the moon.
\end{quotation}
(again, we omit the terms with subscripts $3,4,\dots,n$; we also changed the notation used for the Legendre polynomials to the standard one).
In this equation, $r_{f-\delta}$ is the distance to the Moon when the true anomaly of the Moon is $f-\delta$.

In the several instances in which these equations are used, $f$ is always the true anomaly. 
This can be clearly seen, for instance, in MacDonald's eqn. (50) giving the quantity $C=r_{f-\delta}^{-3}r_f^{-4}$.

\section{Tide heights in MacDonald's Earth}

We may solve the ambiguity of MacDonald's written text without introducing any interpretative hypothesis. 
MacDonald's eqn. (21) is the classical expression of the potential created by a prolate spheroid on an external point (the Moon) and we may solve the inverse problem: 
\begin{quotation}
\noindent
Which is the spheroid creating on the Moon the additional potential given by MacDonald's equation (21) ? 
\end{quotation}

Thereafter, for the sake of avoiding embedding the point that we want to elucidate into a huge amount of algebra, we consider the planar approximation in which the reference plane is the plane of the lunar motion, the origin axis passes through the Moon's orbit perigee and the Earth rotation axis is perpendicular to the Moon's orbital plane. None of these assumptions interferes with the nature of the problem being considered.

The major principal axis of the spheroid may lie on the intersection with the reference (orbital) plane, of one cone whose axis is directed to the Moon and having half-opening $\delta$.
Mathematically we have two solutions, but one of them is on the wrong side (the tide is advancing the Moon's position instead of lagging it).
The height over the sphere of the points of the spheroid creating the potential ${\cal{U}}$ are given by:
\beq
\delta r = \frac{5R_\oplus^4}{12 r_{f-\delta}^3} k_2 (2\cos 2\Psi + 1)
\endeq
where $\Psi$ is the angular distance from the generic points of the prolate spheroid to its major principal axis.
If we consider this point in the equator of the body: $\Psi=\phi-(\omega+f-\delta)$ or $\Psi=\phi-f+\delta$ (since we adopted $\omega=0$), where $\phi$ is the longitudfe of the point. 
The harmonic decomposition of the resulting tidal height at that point is 

\beq\begin{array}{ll}
\delta r = &  \displaystyle  \frac{5R_\oplus^4}{72a^3} k_2 \Big( 4 + 12 \cos(2 \ell - 2 \phi - 2 \delta ) +     12 e \cos(\ell - \delta) + 
   18 e \cos(\ell - 2 \phi - \delta ) - \\ & \nonumber 
   24 e \cos(\ell - 2 \phi - 2 \delta) + 
   24 e \cos(3 \ell - 2 \phi - 2 \delta ) + 
   18 e \cos(3 \ell - 2 \phi - 3 \delta ) + 
 \phantom{\Big(} \\ & \nonumber 
   18 e^2 - 12 e^2 \cos(\delta) +
     9 e^2 \cos(2 \phi) - 
   18 e^2 \cos(2 \phi + \delta ) +
     9 e^2 \cos(2 \phi + 2 \delta ) +   12 e^2 \cos(2 \ell - \delta) +  \phantom{\Big(}
\\ & \nonumber 
   6 e^2 \cos(2 \ell - 2 \delta) + 
   18 e^2 \cos(2 \ell - 2 \phi - \delta ) +
   6 e^2 \cos(2 \ell - 2 \phi - 2 \delta ) -
   54 e^2 \cos(2 \ell - 2 \phi - 3 \delta ) + \phantom{\Big(}
\\ & \nonumber 
   39 e^2 \cos(4 \ell - 2 \phi - 2 \delta ) + 
   54 e^2 \cos(4 \ell - 2 \phi - 3 \delta ) + 
     9 e^2 \cos(4 \ell - 2 \phi - 4 \delta ) \Big)  +{\cal{O}}(e^3).
\end{array}\endeq

Since the Earth is rotating with angular velocity $\Omega$, we may write $\phi=\phi_0+\Omega t$. 
The simple inspection of the above equation shows that in the strict MacDonald model the Fourier decomposition of the height of the tide does not show any simple law relating the lags and the frequencies of the tide components. 
In particular, it is not true that in a MacDonald tide all tide components have the same lag.

The conclusion is that, as MacDonald fixed the vertex of the Earth tide at a constant distance $\delta$ from the Moon, his model leads in a univocally determined way to a trigonometric series in which the components lag on a very particular way.  
Even the simple semi-diurnal component has parts with different lags: $\cos(2 \ell - 2 \phi - 2 \delta ), e^2 \cos(2 \ell - 2\phi - \delta ), e^2 \cos(2 \ell  - 2 \phi - 3 \delta)$ (which have the same frequency, but the lags $2\delta$, $\delta$ and $3\delta$ respectively). The same behavior may be verified in  higher orders of approximation.

\section{MacDonald's potential of the deformed Earth}\label{MacDo}

Another possibility is to consider MacDonald's Eqn. 21 as the additional potential at a point whose distance to the Earth is the distance of the Moon, $r_f$, and whose angular distance to the vertex of the tidally deformed Earth is $\delta$. This means that MacDonald indeed introduced a lag in the true anomaly and not in time.

MacDonald's eqn. (21) can be easily converted into the equation for the potential in a generic point $\tens{P}(\rho,\phi)$:
\beq
{\cal{U}}=  \frac{GmR_\oplus^5}{\rho^3 r_{f-\delta}^3} k_2 P_2(\cos \Psi)
\endeq
\noindent where $\Psi$ is the angular distance from the generic point to the vertex of the tidally deformed Earth. This is a classical equation.

This equation for the potential is the only possible one which is consistent with MacDonald's Eqn. (20) and with the words highlighted after that equation, ``the zenith distance becomes $\delta$",  in which the distance of the Moon to the vertex of the Earth's tidal prolate spheroid is a constant $\delta$.

In the following, for simplicity, we assume the point $\tens{P}(\rho,\phi)$ on the Earth's equator (which again is assumed to coincide with the plane of Moon's orbit). Then $\Psi$ is again the difference between the longitude of the point, $\phi$, and the longitude of the delayed Moon, $\omega+f-\delta$. 

Therefore,
\beq
{\cal{U}}=  \frac{GmR_\oplus^5}{\rho^3 r_{f-\delta}^3} k_2 P_2\big(\cos (\phi-f+\delta)\big)
\endeq
\noindent where, for simplicity, we have again adopted $\omega=0$. This function can be expanded in Fourier series to give:
\beq\begin{array}{ll}
{\cal{U}} & = \displaystyle \frac{GmR_\oplus^5}{16\rho^3 a^3} k_2 \Big( 4 + 12 \cos(2 \ell - 2 \phi - 2 \delta ) 
+     12 e \cos(\ell - \delta) +    18 e \cos(\ell - 2 \phi - \delta ) - \\ & \nonumber 
   24 e \cos(\ell - 2 \phi - 2 \delta) + 
   24 e \cos(3 \ell - 2 \phi - 2 \delta ) + 
   18 e \cos(3 \ell - 2 \phi - 3 \delta ) + 
 \phantom{\Big(} \\ & \nonumber 
   18 e^2 - 12 e^2 \cos(\delta) +
     9 e^2 \cos(2 \phi) - 
   18 e^2 \cos(2 \phi + \delta ) +
     9 e^2 \cos(2 \phi + 2 \delta ) +   12 e^2 \cos(2 \ell - \delta) +  \phantom{\Big(}
\\ & \nonumber 
   6 e^2 \cos(2 \ell - 2 \delta) + 
   18 e^2 \cos(2 \ell - 2 \phi - \delta ) +
   6 e^2 \cos(2 \ell - 2 \phi - 2 \delta ) -
   54 e^2 \cos(2 \ell - 2 \phi - 3 \delta ) + \phantom{\Big(}
\\ & \nonumber 
   39 e^2 \cos(4 \ell - 2 \phi - 2 \delta ) + 
   54 e^2 \cos(4 \ell - 2 \phi - 3 \delta ) + 
     9 e^2 \cos(4 \ell - 2 \phi - 4 \delta ) \Big)  +{\cal{O}}(e^3).
\end{array}\endeq

\noindent 
the periodic part of which is the same as found in Sec. 2. 
The Earth potential in the considered point is a Fourier oscillations packet due to the Moon raised tide.
As discussed in the previous section, the potential due to the strict MacDonald model is formed by oscillations that lag in a way not correlated to their frequencies.

The coincidence of this result with that of the previous section proves that, contrariçly to what can be read in his written text, MacDonald has indeed introduced a lag in the true anomaly and not in time.

\section{The Williams-Efroimsky modified theory}

Williams and Efroimsky (2012) consider that a delay cannot be introduced in the true anomaly, which is not uniform.
In their paper,  the delay is introduced in the mean anomaly, and MacDonald's eqn. (21) is rewritten in a form equivalent to 
\beq
{\cal{U}}=  \frac{GmR_\oplus^5}{r(\ell)^3 r(\ell-\delta)^3} k_2 P_2\Big(\cos f(\ell)-f(\ell-\delta)\Big).
\endeq
what means that they are considering not the action of the Moon when its true anomaly was $f-\delta$, but when its mean anomaly was $\ell-\delta$.

We may perform the same calculations as above, to obtain for one generic point $P(\rho,\phi)$:
\beq\begin{array}{ll}
{\cal{U}} & = \displaystyle \frac{GmR_\oplus^5}{8\rho^3 a^3} k_2 \Big(2 + 6 \cos( 2 \ell - 2 \phi + 2 \delta ) + 
    6 e \cos( \ell + \delta) - 3 e \cos( \ell - 2 \phi + \delta ) +    
 \nonumber
\\ & \nonumber
 21 e \cos( 3 \ell - 2 \phi + 3 \delta) +3 e^2 + 
   9 e^2 \cos( 2 \ell + 2 \delta)  -  15 e^2 \cos( 2 \ell - 2 \phi + 2 \delta ) + 
    \phantom{\Big(}
\\ & \nonumber
   51 e^2 \cos( 4 \ell - 2 \phi + 4 \delta )\Big) +{\cal{O}}(e^3)
\end{array}\endeq

Notwithstanding the much more simple expression, a law relating lags and frequencies is not yet visible. For that sake, one more hypothesis is needed (which is the same operational hypothesis introduced by Mignard(1979) in his concise presentation of Darwin's theory): At a given time, the potential on one point on the Earth is equal to the potential which would be acting on it at a time $\tau$ before the given time, would the Earth be inviscid. This means that we have to introduce not the longitude $\phi$ of the point but $\phi-\Omega\tau$, or $\phi-(\Omega\delta/n)$.
The above expansion then becomes
\beq\begin{array}{ll}
{\cal{U}} & = \displaystyle  \frac{GmR_\oplus^5}{8\rho^3 a^3} k_2 \Big(  2 +    6 \cos( 2 \ell - 2 \phi + 2 \delta(1-\Omega/n) ) +
   6 e \cos( \ell + \delta) - 3 e \cos( \ell - 2\phi + \delta(1-2\Omega/n)  + \nonumber
\\ & \nonumber
   21 e \cos( 3 \ell -  2 \phi + 3 \delta(1-2\Omega/3n) ) +   3 e^2  +  9 e^2 \cos( 2 \ell + 2 \delta) )  - 
   15 e^2 \cos( 2 \ell - 2 \phi+ 2 \delta(1-\Omega/n))+ \phantom{\Big(}
\\ & \nonumber
   51 e^2 \cos( 4 \ell -2 \phi+ 4 \delta(1-2\Omega/4n) )\Big) +{\cal{O}}(e^3)
\end{array}\endeq
\noindent where Darwin's law of lags proportional to frequencies emerges making the new theory consistent with the standard Darwinian theories.
This result corroborates what has been found by Ferraz-Mello et al. (2009): The second-degree expansion of the components of the force obtained with the lag time introduced by Mignard (1979) is equivalent to that obtained with Darwin standard theory when lags are assumed proportional to frequencies. 
The last equation can be written in terms of Mignard's constant time lag $\tau$ via $\delta=n\tau$.

\section{Singer's modification}

Singer (1968) also proposed a modification in MacDonald's model to take into account the nonuniformity of $f$. He used MacDonald's equations, but considered that the distance from the vertex of the Earth tide to the Moon is variable, and used
\beq
\delta=\tau\left(\Omega-\frac{df}{dt}\right). \label{patch}
\endeq
Since $\tau =$ const,  Singer's extension is sometimes considered as a pionneer example of theory with constant time-lag; however, when the patch proposed in eqn. (\ref{patch}) is introduced in MacDonald's equations, the result does not 
correspond to a constant time-lag rheology as do, for instance, the Williams-Efroimsky equations. As in the 
original version of MacDonald, no law exists ruling the formation of the lags in Singer's tide components. 
The correction of the geometric lag nonuniformity entails, correctly, the argument $2(\ell - \phi) + 2 (n - \Omega)\tau$ in the semi-diurnal terms of Singer's $U$. However, in the other tide components the arguments do  not follow any law relating lags and frequencies. For instance, we mention the monthly tide arguments $(3\ell - 2 \phi + 3 n \tau - 3 \Omega\tau)$ and 
$(\ell - 2 \phi + 2 n \tau - \Omega\tau)$, the argument $2 \phi$ (without any lag), etc.
Had Singer's patch been enough to endow MacDonald's model with Darwin's rheology, all terms would have the form $k\ell - k^\prime \phi + (k n  - k^\prime \Omega)\tau$. 

\section{Conclusions}
The figure of the tidally deformed Earth in MacDonald's theory has its symmetry axis at a constant distance from the Moon. In that theory, the fact that the Moon is on an elliptic orbit only affects the ratio of the figure axes. The only variable is the prolateness of the spheroid, which is inversely proportional to the cube of the Earth-Moon distance. As a consequence, MacDonald's equations do not allow to classify his theory according with a rheology. The lags of the harmonic tide components only follow the  geometrical constraints of this spheroid. They do not obey any law and are unrelated to the frequencies of the components. The patch proposed by Singer (1968) corrects the arguments of the terms describing the semi-diurnal tide, but for the remaining terms any law is followed.
In order to get one \textit{constant lag-time} theory, it is not sufficient to correct the nonuniformity of $\delta$, but it is also necessary to introduce Mignard's operational hypothesis (Mignard, 1979) after which the actual potential on one point on the Earth,  at a given time, is equal to the potential which would be acting on it at a time $\tau$ before the given time, would the Earth be inviscid. This operational hypothesis is also adopted in the modified MacDonald theory presented by Williams and Efroimsky (2012) and is instrumental to endows it with Darwin's rheology.

\end{document}